\documentclass[aps,prl,twocolumn,preprintnumbers,showpacs]{revtex4-1}

\usepackage{graphicx}
\usepackage{graphics}
\usepackage{dcolumn}
\usepackage{bm}
\usepackage{epstopdf}
\usepackage{mathrsfs}
\usepackage{amssymb}
\usepackage{amsmath}
\usepackage[pdftex,colorlinks=true,linktocpage=true,linkcolor=blue,citecolor=blue]{hyperref}
\usepackage{xspace}
\usepackage{slashed}
\usepackage{nicefrac}
\usepackage{xcolor}
\usepackage[normalem]{ulem}

\def\del{\partial}

\def\p{{\boldsymbol p}}

\def\k{{\boldsymbol k}}
\def\kh{\hat{\boldsymbol{k}}}

\def\d{\text{d}}

\def\sumint{\hbox{$\sum$}\!\!\!\!\!\!\!\,{\int}}
\def\smallsumint{{\hbox{$\scriptstyle\sum$}}\!\!\!\!\!\,{\int}}

\newcommand{\smallG}{{\scriptscriptstyle{G}}}
\newcommand{\smallS}{{\scriptstyle{S}}}

\newcommand{\beq}{\begin{eqnarray}}
\newcommand{\eeq}{\end{eqnarray}}
\newcommand{\be}{\begin{eqnarray*}}
\newcommand{\ee}{\end{eqnarray*}}
\newcommand{\bqa}{\begin{eqnarray}}
\newcommand{\eqa}{\end{eqnarray}}

\newcommand{\nn}{\nonumber\\ }

\begin{document}

\title{Massless Mode and Positivity Violation in Hot QCD}

\author{Nan Su}
\email{nansu@physik.uni-bielefeld.de}
\affiliation{%
Fakult\"at f\"ur Physik, Universit\"at Bielefeld, Universit\"atsstra{\ss}e 25, 33615 Bielefeld, Germany}

\author{Konrad Tywoniuk}
\email{konrad@ecm.ub.edu}
\affiliation{%
Departament d'Estructura i Constituents de la Mat\`eria 
and Institut de Ci\`encies del Cosmos (IC-CUB),\\
Universitat de Barcelona, Mart\'i i Franqu\`es 1, 08028 Barcelona, Spain}

\preprint{BI-TP 2014/12, ICCUB-14-058}

\date{\today}

\begin{abstract}
We calculate the quark self-energy at one-loop level at high temperature, taking into account contributions from both the (chromo)electric scale $gT$ and the (chromo)magnetic scale $g^2T$. While reproducing standard massive excitations due to the electric scale, we uncover a novel massless excitation ascribable to the magnetic scale. The residue of this massless excitation is nonpositive at all temperatures, which consequently gives rise to positivity violation in the quark spectral functions. This demonstrates the profound impact of confinement effects on thermal quark collective excitations, which manifest genuine long-range correlations in the system.
\end{abstract}


\maketitle

As predicted by asymptotic freedom in quantum chromodynamics (QCD), hadronic matter at sufficiently high temperatures and densities undergoes a transition to a novel state of deconfined matter, known as the quark-gluon plasma (QGP), whose properties are strongly influenced by collective excitations. Creating and studying the QGP is one of the main goals of the ultrarelativistic heavy-ion collision experiments at RHIC and LHC. The properties of the created bulk matter display the feature of a strongly coupled liquid characterized by long-range correlations (see Ref.~\cite{Teaney:2009qa} and references therein). In turn, it is distinguished by a small viscosity, which spoils the expectation from perturbation theory of a weakly coupled plasma where long-range correlations are missing \cite{Arnold:2000dr}. This has generated interest in strong-coupling formalisms, such as the celebrated AdS/CFT correspondence (see Ref.~\cite{CasalderreySolana:2011us} for a review).

Apart from the intrinsic energy scale, the temperature $T$, the collective behavior of a hot QGP gives rise to two additional thermal scales, namely the (chromo)electric scale $gT$ and the (chromo)magnetic scale $g^2T$. While the electric scale can be tackled by resummed perturbation theory (see~\cite{tft} for reviews), the magnetic scale still poses profound challenges. Because of the absence of an infrared (IR) cut off by screening in the magnetic sector, the perturbative expansion of finite-temperature Yang-Mills theory breaks down fundamentally at the magnetic scale, which is the origin of the so-called Linde problem~\cite{linde}. It has been realized that the nonperturbative nature of the magnetic scale is intimately related to the confining properties of the dimensionally reduced Yang-Mills theory at high temperature~\cite{mag}. This strongly suggests that a confinement mechanism has to be incorporated within perturbative resummation even when dealing with the QGP phase.

A formalism to tackle this issue is the Gribov-Zwanziger (GZ) action, which is well known from studies of color confinement~\cite{gz}. It regulates the IR behavior of QCD by fixing the IR residual gauge transformations that remain after applying the Faddeev-Popov procedure. The GZ action is renormalizable, and it thus provides a systematic framework for perturbative calculations (i.e., $g\ll1$) incorporating confinement effects which is the course of study here. The gluon propagator in the general covariant gauge from the GZ action reads
\beq
D^{\mu\nu}(P)
\,=\,
\left[ \delta^{\mu\nu} - (1-\xi)\frac{P^\mu P^\nu}{P^2} \right] \frac{P^2}{P^4 + \gamma_\smallG^4}
\;,
\label{eq:gluon}
\eeq
where $\xi$ is the gauge parameter and $\gamma_\smallG$, called the Gribov parameter, is solved self-consistently from a gap equation that is defined to infinite loop orders (see Ref.~\cite{gribov-review} for reviews). The GZ gluon propagator is IR suppressed, manifesting confinement effects~(see Ref.~\cite{Maas:2011se} for a review), and is a significant improvement over the one arising in the original Faddeev-Popov quantization that forms the basis for perturbative calculations. The gap equation at one-loop order can be solved analytically at asymptotically high temperatures and gives~\cite{Zwanziger:2006sc,Fukushima:2013xsa} 
\beq
\gamma_\smallG = \frac{D-1}{D} \frac{N_c}{4\sqrt{2}\pi} g^2T \,,
\label{eq:gamma}
\eeq
where $D$ is the space-time dimensions and $N_c$ is the number of colors. Equation~(\ref{eq:gamma}) provides a fundamental IR cutoff at the magnetic scale for the GZ action. Because of the incorporation of the magnetic scale, the resulting equations of state show stable and robust behavior that is consistent with lattice data down to nearly the deconfinement transition~\cite{Zwanziger:2004np,Fukushima:2013xsa}, in contrast to results from conventional resummed perturbation theory~\cite{resum}. These results therefore highlight the significant role played by the magnetic scale in the phenomenologically relevant temperature regime.

An important measure of collective behavior in a hot QGP is the self-energy of quarks and gluons, from which screening masses, dispersion relations, and spectral functions of collective excitations are derived. In this context it is worth pointing out the intimate connection between two fundamental features of long-range correlations and confinement in QCD. On one hand, color confinement is deeply related to positivity violation of the spectral function~\cite{Alkofer:2000wg}, which has been intensively studied using lattice QCD and functional methods at both zero and finite temperatures for gluons (see Ref.~\cite{pv-g} and references therein). The study of the quark sector has not been equally conclusive~\cite{pv-q}. Positivity violation is so far a missing feature in resummed perturbation theory calculations. On the other hand, in order to describe a strongly coupled QGP, massless modes that incorporate long-range correlations have been studied in functional methods~\cite{zero-mode} and through the AdS/CFT correspondence~\cite{ads}. While massless modes have been studied using resummed perturbation theory for the Nambu--Jona-Lasinio model, the Yukawa model, QED and perturbative QCD~\cite{ultra}, there has not yet been a proposal for QCD that would render the system strongly coupled.

In this Letter, we present a first study of the thermal quark self-energy incorporating effects from both electric and magnetic scales (via the GZ action at one-loop level) at the high-temperature limit (where $g\ll1$) using systematics of the hard-thermal-loop (HTL) effective theory~\cite{Braaten:1989mz}. There have been similar studies for the quark self-energy with nonperturbative gluons at finite density~\cite{mu} and in strong magnetic fields~\cite{Kojo:2012js}. While reproducing the standard massive excitations due to the electric scale, we uncover a novel massless excitation ascribable to the magnetic scale. This massless mode induces positivity violation of the quark spectral functions and incorporates long-range correlations in the system. We are aware of the fact that the refined Gribov-Zwanziger action is in better agreement with lattice data~\cite{Dudal:2008sp}. However, the emergence of the massless mode is exclusively due to the nature of complex conjugate poles, which is a shared feature of all Gribov-like approaches. We thus use the GZ action as a simple demonstration, without the loss of generality.

The Euclidean one-loop quark self-energy reads
\beq
\Sigma(P)
\,=\,
(ig)^2 C_F \sumint_{\!\!\!\!\{ K \}} \gamma^\mu S(K) \gamma^\nu D^{\mu\nu}(P-K)
\;,
\label{eq:Sigma}
\eeq
where $g$ is the running coupling, $C_F= (N_c^2-1)/2N_c$ is the quadratic Casimir operator (in fundamental representation), $S(P)=1/{/\!\!\!\!P}$ is the quark propagator, and $D^{\mu\nu}(P)$ is the gluon propagator, which is taken from Eq.~(\ref{eq:gluon}) in our calculation~\cite{fn-1}.

It is well known that the gauge-dependent terms in the one-loop quark self-energy do not contribute at the mass dimension of $[T^2]$, which is relevant for the screening mass, dispersion relations and spectral functions~\cite{mq}, and thus the appropriate gauge-invariant contribution [equivalent to setting $\xi = 0$ in Eq.~(\ref{eq:Sigma})] for our purposes after carrying out the Matsubara summation reads
\begin{align}
&\Sigma(P) = (ig)^2 C_F \frac{2-D}{2} \sum_\pm \int \frac{\d^3k}{(2\pi)^3} \frac{1}{4E_\pm} \Bigg\{ \big[1 +  n_B(E_\pm) \nonumber
\\ &-n_F(k) \big] \left[ \frac{i\gamma_0 + \kh\cdot{\boldsymbol\gamma}}{iP_0 + k + E_\pm} + \frac{i\gamma_0 - \kh\cdot{\boldsymbol\gamma}}{iP_0 - k - E_\pm} \right] + \big[n_B(E_\pm) \nonumber
\\  &+ n_F(k) \big] \left[\frac{i\gamma_0 + \kh\cdot{\boldsymbol\gamma}}{iP_0 + k - E_\pm} + \frac{i\gamma_0 - \kh\cdot{\boldsymbol\gamma}}{iP_0 - k + E_\pm} \right] \Bigg\} \,,
\label{eq:Sigma1}
\end{align}
where $E_\pm=\sqrt{(\k-\p)^2 \pm i\gamma_\smallG^2}$, $n_B$ and $n_F$ are the Bose-Einstein and Fermi-Dirac distributions, and $\kh = \k/k$ and $k= |\k|$ is the norm of the three-dimensional vector. In the following, we set the space-time dimension $D=4$.

\begin{figure}[t!]
\centering
\includegraphics[width=0.45\textwidth]{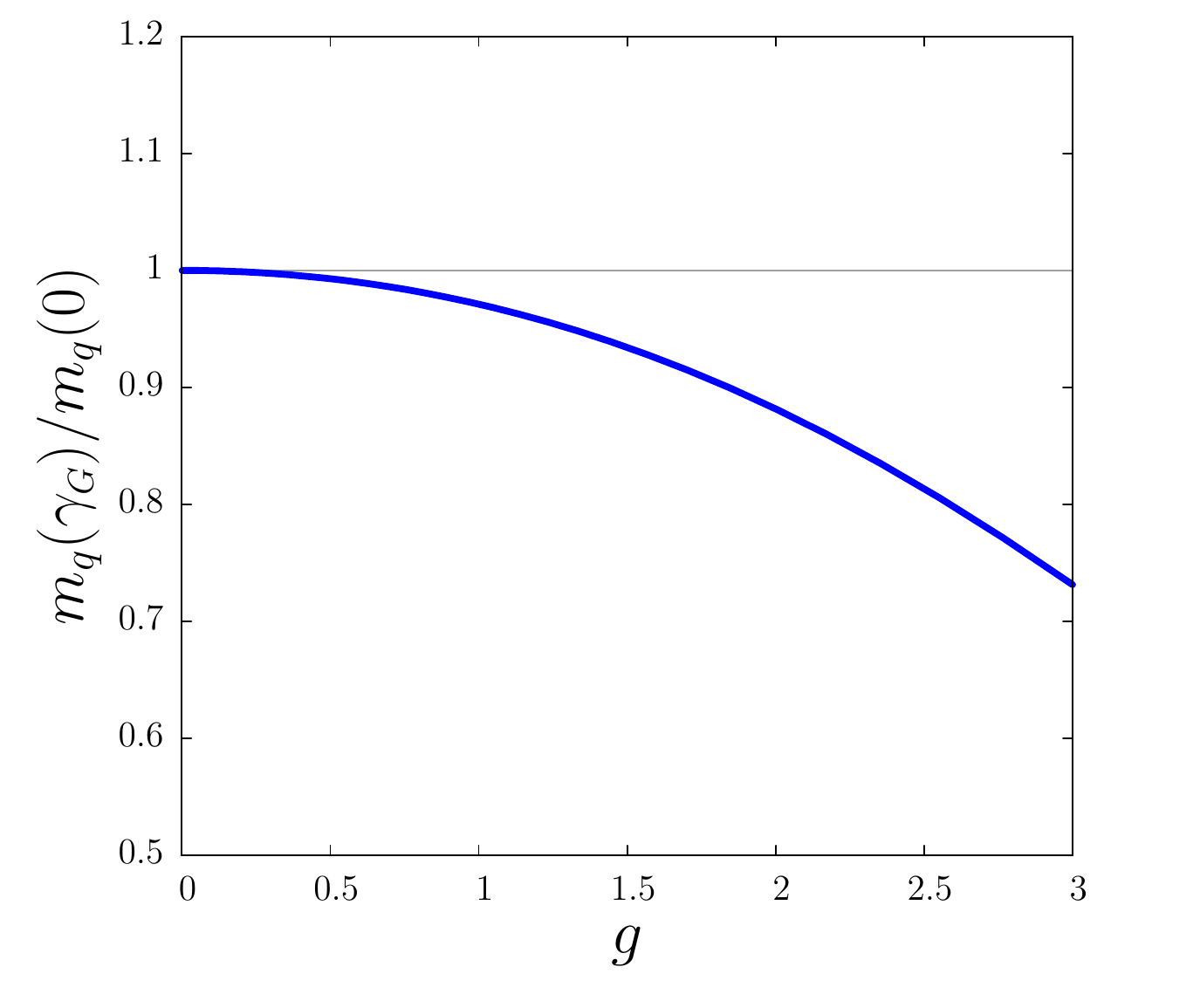}
\caption{\label{fig:mq} The quark screening mass $m_q(\gamma_\smallG)$ from Eq.~(\ref{eq:mq1}), normalized by the perturbative value $m_q(0)$.
}
\end{figure}

From the HTL effective theory, we know that the leading contribution to the two-point correlation function  from $\Sigma(P)$ stems from soft external momenta $P \ll K$~\cite{Braaten:1989mz}. For this calculation we are interested in the high-$T$ behavior of the self-energy in the small-$g$ regime; therefore, we may expand Eq.~(\ref{eq:Sigma1}) in terms of small $P=(P_0,\p)$ over $\k$ by using $n_B(E_\pm) \simeq n_B(E_\pm^0)$, $k + E_\pm \simeq 2k$, and $k - E_\pm \simeq k - E_\pm^0 + \p\cdot\k \big/E_\pm^0$, where $E_\pm^0=\sqrt{k^2 \pm i\gamma_\smallG^2}$. As a result, Eq.~(\ref{eq:Sigma1}) becomes
\begin{align}
\label{eq:Sigma2}
&\Sigma(P) \simeq -(ig)^2 C_F 
\sum_\pm \int_0^\infty\frac{\d k}{2\pi^2}k^2 \int\frac{\d\Omega}{4\pi} \frac{\tilde n_\pm(k,\gamma_\smallG)}{4E_\pm^0} \\ 
& \hspace{1em}\times
\left[
\frac{i\gamma_0 + \kh \cdot{\boldsymbol\gamma}}{iP_0 + k - E_\pm^0 + \frac{ \p \cdot \k}{E_\pm^0}}
+
\frac{i\gamma_0 - \kh \cdot{\boldsymbol\gamma}}{iP_0 - k + E_\pm^0 - \frac{ \p \cdot \k}{E_\pm^0}}
\right]
\,, \nonumber
\end{align}
where $\tilde n_\pm(k,\gamma_\smallG)\equiv n_B(\sqrt{k^2 \pm i \gamma_\smallG^2}) + n_F(k)$ and $\int \d\Omega = \int_0^{2\pi}\d\phi \int_0^\pi \d\cos\theta$ denotes the measure for the angular integral. The quark screening mass $m_q$ squared is conventionally defined as the coefficient in front of the angular integral of the self-energy in the HTL setup. By analogy, the quark screening mass, which in our calculation incorporates effects from the magnetic scale, reads
\beq
m_q^2(\gamma_\smallG) = g^2 C_F 
\frac{1}{4\pi^2} 
\sum_\pm \int_0^\infty dk \frac{k^2}{E_\pm^0} \tilde n_\pm(k,\gamma_\smallG) \,,
\label{eq:mq1}
\eeq
which reduces to the conventional quark screening mass $m_q^2(0) =C_F g^2T^2/8$ for $\gamma_\smallG= 0$. In this exploratory study, we will neglect running coupling effects. In Fig.~\ref{fig:mq} we show the resulting $m_q(\gamma_\smallG)$ from Eq.~(\ref{eq:mq1}) normalized by its value from conventional resummed perturbation theory. It is clear from the plot that $m_q$ receives negative contributions from $\gamma_\smallG$, which is a manifestation of antiscreening effects generated by the magnetic scale. Although the effect is modest in the studied range of the coupling, this is a profound signal of the buildup of long-range correlations in the system and similar antiscreening effects have been observed on the lattice for the Debye screening mass~\cite{Kaczmarek:2005ui}.

The quark screening mass in Eq.~(\ref{eq:mq1}) also plays a crucial role for the active fermionic degrees of freedom that emerge from studying the dispersion relation. In order to calculate properties of the real-time retarded propagator we need to perform an analytical continuation to Minkowski space.
\begin{figure}[t]
\centering
\includegraphics[width=0.5\textwidth]{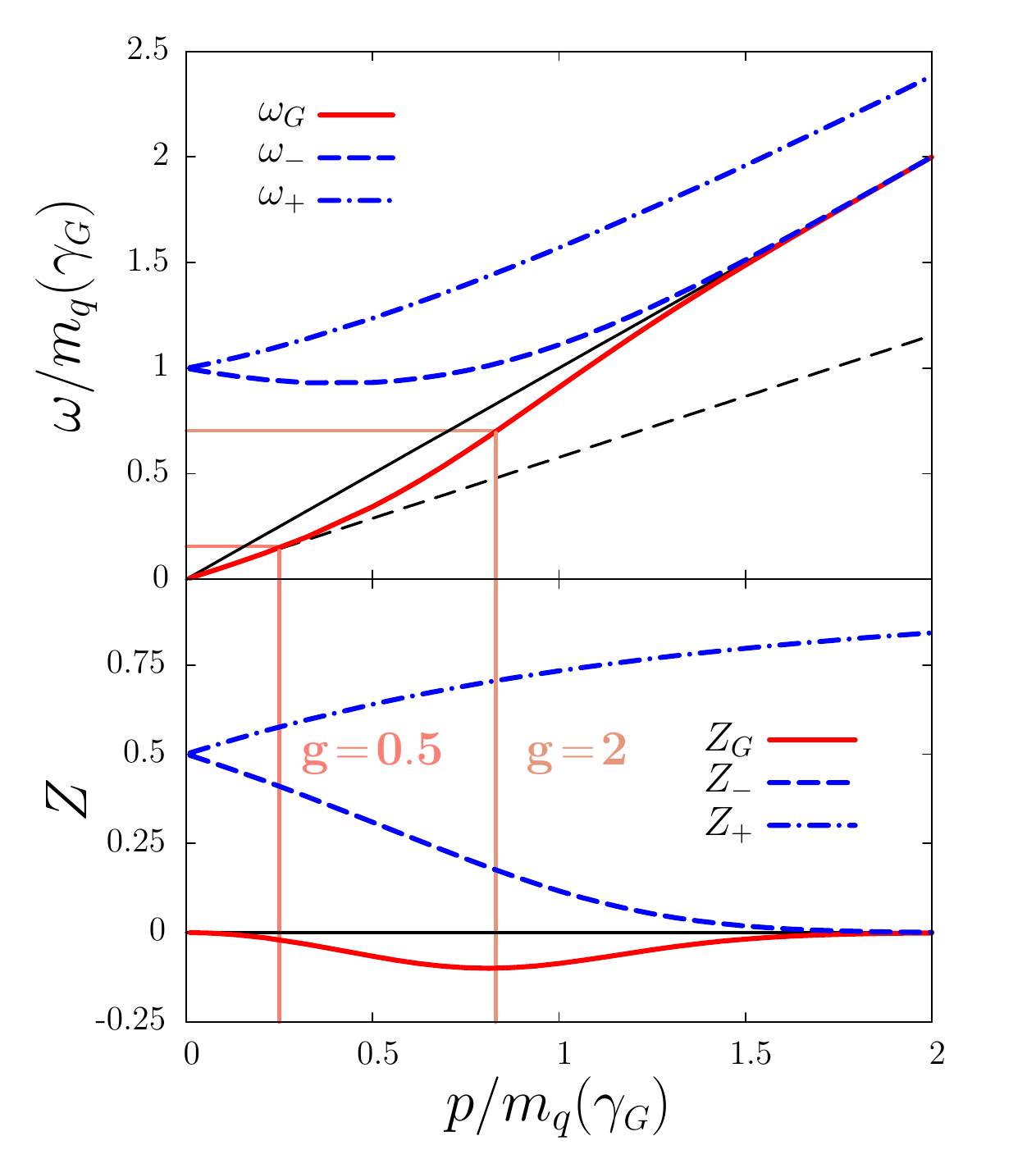}
\caption{\label{fig:DR&Res} 
Dispersion relations (upper panel) and the corresponding residues (lower panel) for the quark ($\omega_+$), antiquark or plasmino ($\omega_-$), and Gribov ($\omega_\smallG$) poles. The scaling with $m_q(\gamma_\smallG)$ has been numerically established for $g \lesssim 2$.}
\end{figure}
We define the Minkowskian resummed quark propagator
\beq
\label{eq:ResumProp}
iS^{-1}(P) = \slashed P - \Sigma(P) = A_0 \gamma_0 - A_\smallS \hat \p\cdot {\boldsymbol\gamma} \,,
\eeq
which can be decomposed into positive and negative helicity-to-chirality contributions, as usual \cite{LeBellac:2004}. Defining the modified frequencies 
$\tilde\omega^\pm_1 \equiv E_\pm^0(\omega + k - E_\pm^0)/k$ 
and 
$\tilde\omega^\pm_2 \equiv E_\pm^0(\omega - k + E_\pm^0)/k$, 
the angular integrals follow straightforwardly by making use of the Legendre functions $Q_0(\omega,p) \equiv \frac{1}{2 p}\log\frac{\omega + p}{\omega - p}$ and $Q_1(\omega,p)\equiv [1-\omega Q_0(\omega, p)]/p$. With the help of these definitions the coefficients $A_0$ and $A_\smallS$ in Eq.~(\ref{eq:ResumProp}) can be written as
\begin{align}
\label{eq:A0}
A_0(\omega,p) &= \omega - \frac{2g^2C_F}{(2\pi)^2}\sum_{\pm}\int\d k\, k\,\tilde n_\pm(k,\gamma_\smallG) \nn 
&\;\;\;\;\;\times \left[Q_0(\tilde\omega^\pm_1,p) + Q_0(\tilde\omega^\pm_2,p) \right] \,,\\
\label{eq:AS}
A_\smallS(\omega,p) &= p + \frac{2g^2C_F}{(2\pi)^2} \sum_\pm\int\d k \,k\, \tilde n_\pm(k,\gamma_\smallG) \nn 
&\;\;\;\;\;\times \left[Q_1(\tilde\omega^\pm_1,p) + Q_1(\tilde\omega^\pm_2,p) \right] \,.
\end{align}
Because of parity symmetry \cite{LeBellac:2004}, it is sufficient to study the singularities of the denominator $\Delta_+ (\gamma_\smallG)\equiv [A_0(p_0,p) - A_\smallS(p_0,p)]^{-1}$ and their residues, defined as $Z^{-1}\equiv \del \Delta(p_0,p)^{-1}/\del p_0 |_{p_0 = \omega_\text{pole}}$, to map out the pole structure of the full propagator in Eq.~(\ref{eq:ResumProp}). The structures of singularities and branch cuts of Eq.~(\ref{eq:ResumProp}) are more involved than in the conventional case, so the Wick rotation $iP_0 \to \omega + i0^+$ should be treated with more care. In the conventional HTL effective theory, there is a branch cut in the self-energy spanning $\omega = \pm p$. This forbids a Wick rotation in the spacelike regime, $\omega < p$, and consequently prohibits the existence of any collective excitations there. Because of the complex-conjugate structures in Eqs.~(\ref{eq:A0}) and (\ref{eq:AS}), the branch cuts are pushed away from the origin of the complex $\omega$ plane, which sanctions the use of the Wick rotation there. The radius of the allowed regime grows with the scale $g^2T$, which effectively reveals a novel \emph{magnetic scaling} for non-Abelian plasmas. The analytic-continuation property of our setup in the timelike regime, $\omega > p$, is the same as in the conventional HTL effective theory. Further details will be provided in an upcoming publication.

In Fig.~\ref{fig:DR&Res} we plot the dispersion relations (upper panel) and their corresponding residues (lower panel) as a function of the momentum scaled by the quark screening mass defined in Eq.~(\ref{eq:mq1}). In contrast to the conventional HTL expectation, we find three poles in the propagator. First of all, we recover the screened excitations of the plasma, $\omega = \omega_+(p; \gamma_\smallG)$ and $\omega = \omega_-(p;\gamma_\smallG)$, the so-called quark and antiquark or plasmino quasiparticles depicted by dashed-dotted and dashed (blue) curves in Fig.~\ref{fig:DR&Res}, respectively. As expected, we confirm numerically that the low-momentum limit of the dispersion relation corresponds to the screening mass, $\omega_\pm (0; \gamma_\smallG) = m_q(\gamma_\smallG)$. This demonstrates that our definition of the screening mass captures the main modification of these modes for a wide range of couplings. The familiar behavior of the corresponding residues $Z_\pm(p)$ also follows the same expectations, see the bottom panel of Fig.~\ref{fig:DR&Res}. Although the calculation is carried out at $g\ll1$, the results are independent of $g$ and we have verified this explicitly up to $g\sim2$, which covers the phenomenologically most relevant regimes. This $g$-independent behavior is in direct analogy to the conventional HTL case~\cite{LeBellac:2004}.

Surprisingly, we uncover a novel excitation that we call the Gribov pole, $\omega = \omega_\smallG(p; \gamma_\smallG)$. It describes \emph{massless} fermionic excitations in the plasma that are governed by the dispersion relation $\omega = v_s p$ at small momenta, see the solid (red) curves in Fig.~\ref{fig:DR&Res}. We have numerically established $v_s \approx 1/\sqrt{3}$ (speed of sound) which is independent of the coupling for the studied range. According to our discussion above, the allowed range for the Gribov mode ``grows'' in the $(\omega, p)$ plane with increasing $g$ due to the magnetic scaling. The vertical lines in Fig.~\ref{fig:DR&Res} delimit the range of permitted momenta for the Gribov mode according to the {\it magnetic scaling} for two values of $g$. This behavior indicates that the massless mode plays a bigger role as $T$ is decreased towards the deconfinement transition, see also Ref.~\cite{Chernodub:2007rn}. For momenta larger than the permitted one for a given coupling, we enter the regime of branch cuts and Landau damping takes place.

Moreover, the pole goes along with a residue $Z_\smallG(p) < 0$, which directly implies positivity violation of the corresponding spectral functions in the region of spacelike momenta. This implies that this excitation does not possess a K{\"a}llen-Lehmann representation~\cite{KapustaGale:2006} and, accordingly, does not represent a physical quasiparticle excitation and signals effects of confinement~\cite{Alkofer:2000wg}. In our case it can ultimately be traced back to the fact that the structure of Eqs.~(\ref{eq:A0}) and (\ref{eq:AS}) always brings about canceling imaginary parts \cite{Gribov:1999ui} (theories with similar properties have been studied in the context of $i$-particles~\cite{Baulieu:2009ha}). Consequently, the spacelike regime, which is attributed to Landau damping in the conventional HTL effective theory, also accommodates collective excitations. The residue as a function of the rescaled momentum, $Z_\smallG[p/m_q(\gamma_\smallG)]$, is also scaling with $m_q$ and tends to zero at vanishing momenta, $Z_\smallG(0) = 0$. These novel features are direct manifestations of the long-range confinement effects surviving at finite temperatures in the plasma. This is to our knowledge the first result showing profound relations between the massless mode and positivity violation.

In conclusion, we have calculated the properties of fermionic excitations in a hot QGP. Our setup, which incorporates the nonperturbative magnetic screening scale via the Gribov-Zwanziger action, has revealed a novel massless excitation in addition to the conventional quasiparticle and hole excitations. This mode induces positivity violation in the quark spectral functions, which falls in line with expectations from lattice QCD and functional methods. For a wide range of couplings---including the most relevant for the QGP created at RHIC and LHC energies, $g \lesssim 2$---the characteristics of these modes are controlled by a single function of the screening mass $m_q(\gamma_\smallG)$, which is in direct analogy to the conventional HTL case. In addition, the emergent magnetic scaling protects the existence of the massless mode. It is a generic feature of Gribov-like approaches that gluon propagators possess complex-conjugate poles~\cite{gz,Dudal:2008sp} (see also Refs.~\cite{gribov-review,Maas:2011se} for reviews). This property gives rise to the complex-conjugate structures in deriving the coefficients $A_0$ and $A_\smallS$ in the resummed quark propagator~(\ref{eq:ResumProp}), and consequently generates the massless mode. Therefore, the results presented in this Letter reflect common features of Gribov-like approaches, though the calculation is concretely done using the GZ action as a simple demonstration.

The obtained results in our setup are genuine non-Abelian effects and rely on the importance of the magnetic scale, which in the weak-coupling regime governs the dynamics of large distances. Our results have shed new light on the active degrees of freedom released in the course of a heavy-ion collision and will have profound effects on phenomenological interpretations of experimental data. A recent shear viscosity calculation in Yang-Mills theory using positivity-violating spectral functions has obtained encouraging first results in line with expectations of a strongly coupled plasma \cite{Haas:2013hpa}. Furthermore, a massless mode in the plasma can be a source of Cherenkov radiation \cite{Ruppert:2005uz} and can give rise to instabilities in the thermodynamic quantities.

\section*{Acknowledgments}
We acknowledge N.~Borghini, J.~Casalderrey-Solana, M.~Chernodub, D.~Dudal, M.~S.~Guimaraes, O.~Kaczmarek, B.~W.~Mintz, L.~F.~Palhares, J.~M.~Pawlowski, and M.~Strickland for stimulating discussions. N.S. was supported by the Bielefeld Young Researchers' Fund. K.T. was supported by a Juan de la Cierva fellowship and by the research Grants No. FPA2010-20807 and No. 2009SGR502, by the Consolider CPAN project, and by FEDER.


\begin{thebibliography}{99}

\bibitem{Teaney:2009qa} 
  D.~A.~Teaney,
  arXiv:0905.2433 [nucl-th].
  
\bibitem{Arnold:2000dr} 
  P.~B.~Arnold, G.~D.~Moore, and L.~G.~Yaffe,
  {J.\ High Energy Phys.\ } 11 (2000) 001.

\bibitem{CasalderreySolana:2011us}
  J.~Casalderrey-Solana, H.~Liu, D.~Mateos, K.~Rajagopal, and U.~A.~Wiedemann,
  arXiv:1101.0618 [hep-th].

\bibitem{tft}  
  J.-P.~Blaizot, E.~Iancu, and A.~Rebhan,
  hep-ph/0303185;
  U.~Kraemmer and A.~Rebhan,
  {Rep.\ Prog.\ Phys.\ } {\bf 67}, 351 (2004);
  J.~O.~Andersen and M.~Strickland,
  {Ann.\ Phys.\ (N.Y.)} {\bf 317}, 281 (2005);
  N.~Su,
  {Commun.\ Theor.\ Phys.\ } {\bf 57}, 409 (2012).
  
\bibitem{linde}
  A.~D.~Linde,
  {Phys.\ Lett.\ } {\bf 96B}, 289 (1980);
  D.~J.~Gross, R.~D.~Pisarski, and L.~G.~Yaffe,
  {Rev.\ Mod.\ Phys.\ } {\bf 53}, 43 (1981).

\bibitem{mag}
  W.~Buchm\"uller and O.~Philipsen,
  {Nucl.\ Phys.\ } {\bf B443}, 47 (1995);
  G.~Alexanian and V.~P.~Nair,
  {Phys.\ Lett.\ B} {\bf 352}, 435 (1995);
  R.~Jackiw and S.~-Y.~Pi,
  {Phys.\ Lett.\ B} {\bf 368}, 131 (1996);
  {Phys.\ Lett.\ B} {\bf 403}, 297 (1997);
  J.~M.~Cornwall,
  {Phys.\ Rev.\ D} {\bf 57}, 3694 (1998);
  F.~Eberlein,
  {Phys.\ Lett.\ B} {\bf 439}, 130 (1998);
  D.~Bieletzki, K.~Lessmeier, O.~Philipsen, and Y.~Schr\"oder,
  {J.\ High Energy Phys.\ } 05 (2012) 058.

\bibitem{gz}
  V.~N.~Gribov,
  {Nucl.\ Phys.\ } {\bf B139}, 1 (1978);
  D.~Zwanziger,
  {Nucl.\ Phys.\ } {\bf B323}, 513 (1989).

\bibitem{gribov-review}
  Y.~L.~Dokshitzer and D.~E.~Kharzeev,
  {Annu.\ Rev.\ Nucl.\ Part.\ Sci.\ } {\bf 54}, 487 (2004);
  N.~Vandersickel and D.~Zwanziger,
  {Phys.\ Rep.\ } {\bf 520}, 175 (2012).
  
\bibitem{Maas:2011se}
  A.~Maas,
  {Phys.\ Rep.\ } {\bf 524}, 203 (2013).
  
\bibitem{Zwanziger:2006sc}
  D.~Zwanziger,
  {Phys.\ Rev.\ D} {\bf 76}, 125014 (2007).
  
\bibitem{Fukushima:2013xsa}
  K.~Fukushima and N.~Su,
  {Phys.\ Rev.\ D} {\bf 88}, 076008 (2013).
  
\bibitem{Zwanziger:2004np}
  D.~Zwanziger,
  {Phys.\ Rev.\ Lett.\ } {\bf 94}, 182301 (2005).

\bibitem{resum}
  J.-P.~Blaizot, E.~Iancu, and A.~Rebhan,
  {Phys.\ Rev.\ D} {\bf 68}, 025011 (2003);
  A.~Hietanen, K.~Kajantie, M.~Laine, K.~Rummukainen, and Y.~Schr\"oder,
  {Phys.\ Rev.\ D} {\bf 79}, 045018 (2009);
  J.~O.~Andersen, M.~Strickland, and N.~Su,
  {Phys.\ Rev.\ Lett.\ } {\bf 104}, 122003 (2010);
  J.~O.~Andersen, L.~E.~Leganger, M.~Strickland, and N.~Su,
  {Phys.\ Lett.\ B} {\bf 696}, 468 (2011).
  
\bibitem{Alkofer:2000wg}
  R.~Alkofer and L.~von Smekal,
  {Phys.\ Rep.\ } {\bf 353}, 281 (2001).

\bibitem{pv-g}
  A.~Cucchieri, T.~Mendes and A.~R.~Taurines,
  {Phys.\ Rev.\ D} {\bf 71}, 051902 (2005);
  A.~Sternbeck, E.-M.~Ilgenfritz, M.~M\"uller-Preussker, A.~Schiller, and I.~L.~Bogolubsky,
  {\it Proc.\ Sci.,\ } LAT2006 (2006) 076;
  P.~O.~Bowman, U.~M.~Heller, D.~B.~Leinweber, M.~B.~Parappilly, A.~Sternbeck, L.~von Smekal, A.~G.~Williams, and J.~Zhang,
  {Phys.\ Rev.\ D} {\bf 76}, 094505 (2007);
  P.~J.~Silva, O.~Oliveira, D.~Dudal, P.~Bicudo, and N.~Cardoso,
   {\it Proc.\ Sci.,\ } QCD-TNT-III2014 (2014) 040;
  R.~Alkofer, W.~Detmold, C.~S.~Fischer, and P.~Maris,
  {Phys.\ Rev.\ D} {\bf 70}, 014014 (2004);
  A.~Maas, J.~Wambach, B.~Gruter, and R.~Alkofer,
  {Eur.\ Phys.\ J.\ C} {\bf 37}, 335 (2004);
  A.~Maas, J.~Wambach, and R.~Alkofer,
  {Eur.\ Phys.\ J.\ C} {\bf 42}, 93 (2005);
  C.~S.~Fischer, A.~Maas and, J.~M.~Pawlowski,
  {Ann.\ Phys.\ (N.Y.)} {\bf 324}, 2408 (2009);
  L.~Fister and J.~M.~Pawlowski,
  arXiv:1112.5440 [hep-ph];
  S.~Strauss, C.~S.~Fischer, and C.~Kellermann,
  {Phys.\ Rev.\ Lett.\ } {\bf 109}, 252001 (2012).
  
\bibitem{pv-q}
  F.~Karsch and M.~Kitazawa,
  {Phys.\ Lett.\ B} {\bf 658}, 45 (2007);
  {Phys.\ Rev.\ D} {\bf 80}, 056001 (2009);
  O.~Kaczmarek, F.~Karsch, M.~Kitazawa, and W.~S\"oldner,
  {Phys.\ Rev.\ D} {\bf 86}, 036006 (2012).
  J.~A.~Mueller, C.~S.~Fischer, and D.~Nickel,
  {Eur.\ Phys.\ J.\ C} {\bf 70}, 1037 (2010);
  S.-X.~Qin and D.~H.~Rischke,
  {Phys.\ Rev.\ D} {\bf 88}, 056007 (2013).
  
\bibitem{zero-mode}
  S.-X.~Qin, L.~Chang, Y.-X.~Liu, and C.~D.~Roberts,
  {Phys.\ Rev.\ D} {\bf 84}, 014017 (2011);
  F.~Gao, S.-X.~Qin, Y.-X.~Liu, C.~D.~Roberts, and S.~M.~Schmidt,
  {Phys.\ Rev.\ D} {\bf 89}, 076009 (2014).

\bibitem{ads}
  G.~Policastro, D.~T.~Son, and A.~O.~Starinets,
  {J.\ High Energy Phys.\ } 12 (2002) 054;
  P.~K.~Kovtun and A.~O.~Starinets,
  {Phys.\ Rev.\ D} {\bf 72}, 086009 (2005).

\bibitem{ultra}
  M.~Kitazawa, T.~Kunihiro, and Y.~Nemoto,
  {Phys.\ Lett.\ B} {\bf 633}, 269 (2006);
  M.~Kitazawa, T.~Kunihiro, and Y.~Nemoto,
  {Prog.\ Theor.\ Phys.\ } {\bf 117}, 103 (2007);
  M.~Kitazawa, T.~Kunihiro, K.~Mitsutani, and Y.~Nemoto,
  {Phys.\ Rev.\ D} {\bf 77}, 045034 (2008);
  Y.~Hidaka, D.~Satow, and T.~Kunihiro,
  {Nucl.\ Phys.\ } {\bf A876}, 93 (2012);
  D.~Satow,
  {Phys.\ Rev.\ D} {\bf 87}, 096011 (2013);
  J.-P.~Blaizot and D.~Satow,
  {Phys.\ Rev.\ D} {\bf 89}, 096001 (2014).
  
\bibitem{Braaten:1989mz}
  E.~Braaten and R.~D.~Pisarski,
  {Nucl.\ Phys.\ } {\bf B337}, 569 (1990).
  
\bibitem{mu}  
  T.~Kojo, Y.~Hidaka, L.~McLerran, and R.~D.~Pisarski,
  {Nucl.\ Phys.\ } {\bf A843}, 37 (2010);
  T.~Kojo, Y.~Hidaka, K.~Fukushima, L.~D.~McLerran, and R.~D.~Pisarski,
  {Nucl.\ Phys.\ } {\bf A875}, 94 (2012).
  
\bibitem{Kojo:2012js}
  T.~Kojo and N.~Su,
  {Phys.\ Lett.\ B} {\bf 720}, 192 (2013).
  
\bibitem{Dudal:2008sp}
  D.~Dudal, J.~A.~Gracey, S.~P.~Sorella, N.~Vandersickel, and H.~Verschelde,
  {Phys.\ Rev.\ D} {\bf 78}, 065047 (2008).

\bibitem{fn-1}
  Since we do not deal with renormalization issues in the current work, the sum-integral is given simply by $\smallsumint_{P/\{P\}} \equiv T\sum_{P_0} \int \d^{3}\p/(2 \pi)^3$, where for bosonic momentum $P = (2n\pi T, \p)$ and for fermionic momentum $\{P\}=(2n\pi T+\pi T,\p)$.

\bibitem{mq}  
  V.~V.~Klimov,
  {Zh.\ Eksp.\ Teor.\ Fiz.\ }  {\bf 82}, 336 (1982)
  [{Sov.\ Phys.\ JETP} {\bf 55}, 199 (1982)];
  H.~A.~Weldon,
  {Phys.\ Rev.\ D} {\bf 26}, 2789 (1982).

\bibitem{Kaczmarek:2005ui} 
  O.~Kaczmarek and F.~Zantow,
  {Phys.\ Rev.\ D} {\bf 71}, 114510 (2005).

\bibitem{LeBellac:2004}
M.~Le Bellac, {\it Thermal Field Theory} (Cambridge University Press, Cambridge, England, 1996).

\bibitem{Chernodub:2007rn} 
  M.~N.~Chernodub and V.~I.~Zakharov,
  {Phys.\ Rev.\ Lett.\ } {\bf 100}, 222001 (2008).

\bibitem{KapustaGale:2006}
J.~I.~Kapusta and C.~Gale, {\it Finite-Temperature Field Theory: Principles and Applications} (Cambridge University Press, Cambridge, England, 2006).

\bibitem{Baulieu:2009ha}
  L.~Baulieu, D.~Dudal, M.~S.~Guimaraes, M.~Q.~Huber, S.~P.~Sorella, N.~Vandersickel, and D.~Zwanziger,
  {Phys.\ Rev.\ D} {\bf 82}, 025021 (2010).

\bibitem{Gribov:1999ui} 
  V.~N.~Gribov,
  {Eur.\ Phys.\ J.\ C} {\bf 10}, 91 (1999).

\bibitem{Haas:2013hpa}
  M.~Haas, L.~Fister, and J.~M.~Pawlowski,
  {Phys.\ Rev.\ D} {\bf 90}, 091501 (2014).
  
\bibitem{Ruppert:2005uz} 
  J.~Ruppert and B.~M\"uller,
  {Phys.\ Lett.\ B} {\bf 618}, 123 (2005).
  



  
  
  
  
\end{thebibliography}

\end{document}